\documentclass[reprint,
superscriptaddress,
 amsmath,amssymb,
prb,
]{revtex4-2}

\usepackage{graphicx}
\usepackage{dcolumn}
\usepackage{bm}
\usepackage{comment}
\usepackage{xcolor}
\usepackage{MnSymbol}

\begin{document}

\preprint{APS/123-QED}

\title{Local probe investigation of the spin dynamics in the kagome and inter-layers of orthorhombic barlowite Cu$_4$(OD)$_6$FBr: $^{79}$Br and $^{63}$Cu  NQR study}

\author{Takashi Imai}\email{imai@mcmaster.ca}
\affiliation{%
 Department of Physics and Astronomy, McMaster University, Hamilton, Ontario, L8S 4M1, Canada
}%

\author{Jiaming Wang}
\affiliation{%
 Department of Physics and Astronomy, McMaster University, Hamilton, Ontario, L8S 4M1, Canada
}%

 \author{Rebecca W. Smaha}
   \altaffiliation[Present Address: ]{Materials Science Center, National Renewable Energy Laboratory, Golden, CO 80401, USA.}
 \affiliation{Stanford Institute for Materials and Energy Sciences, SLAC National Accelerator Laboratory, Menlo Park, CA 94025, USA}
 \affiliation{Department of Chemistry, Stanford University, Stanford, CA 94305, USA}

\author{Wei He}
\affiliation{Stanford Institute for Materials and Energy Sciences, SLAC National Accelerator Laboratory, Menlo Park, CA 94025, USA}
\affiliation{Department of Materials Science and Engineering, Stanford University,Stanford, CA 94305, USA}

\author{Jiajia Wen}
\affiliation{Stanford Institute for Materials and Energy Sciences, SLAC National Accelerator Laboratory, Menlo Park, CA 94025, USA}

\author{Young S. Lee}
\affiliation{Stanford Institute for Materials and Energy Sciences, SLAC National Accelerator Laboratory, Menlo Park, CA 94025, USA}
\affiliation{Department of Applied Physics, Stanford University, Stanford, CA 94305, USA}


\date{\today}

\begin{abstract}

{We report $^{79}$Br and $^{63}$Cu nuclear quadrupole resonance (NQR) in the paramagnetic state above $T_\text{N} = 15$~K of the antiferromagnetic orthorhombic phase of barlowite Cu$_4$(OD)$_6$FBr consisting of a layered kagome structure.  The divergent behavior of the longitudinal $^{79}(1/T_{1})$ and transverse $^{79}(1/T_{2})$ relaxation rates observed at $^{79}$Br sites evidences that critical slowing down of Cu spin fluctuations sets in below $\sim20$~K.  This means that one or more Cu sites, most likely at the interlayer Cu(3,4,5) sites between the kagome planes, undergo the antiferromagnetic phase transition in a fairly conventional way.  On the other hand, the $^{63}$Cu NQR signal intensity is gradually wiped out below $\sim30$~K, pointing toward gradual spin freezing of the kagome layers instead.  These contrasting findings suggest significant roles played by magnetic frustration effects within the kagome layers.
}

\end{abstract}

\maketitle


\section{Introduction}
The two-dimensional kagome lattice Heisenberg antiferromagnet is formed by a network of corner sharing triangles, and believed to be one of the most promising avenues to realize quantum spin liquids~\cite{Balents2010,Broholm2020,Savary_2017}.  Theoretically, the kagome lattice has many competing ground states with nearly identical energy, and hence its physical properties are sensitive to structural disorders~\cite{Kawamura2014,Kawamura2019}, spin vacancies within the kagome plane~\cite{Singh2010}, and possible coupling with extra defect spins outside the kagome planes~\cite{Kimchi2018}.  

Among the most intensively studied kagome material lately is (Zn$_{1-x}$Cu$_{x}$)Cu$_3$(OH)$_{6}$FBr ($0.05 \lesssim x \leqq 1$) with kagome layers formed by corner sharing triangles of Cu$^{2+}$ ions with spin-1/2~\cite{Feng2017,SMAHA2018123,Feng2018barlowite,Tustain:2020aa,Smaha2020,Smaha2020_PRM,Tustain:2020aa,Wei_2020,Fu2021,WangNatPhys2021,WangPRL2022,Yuan2022,SmahaRIXSprb}.  The Cu-Cu super exchange interaction $J$ across the O$^{2-}$ sites within the kagome plane is estimated as $J \sim 160$~K~\cite{Jeschke2015}.  These kagome layers are separated by interlayers consisting of Zn$^{2+}$, Cu$^{2+}$, F$^{-}$, and Br$^{-}$ ions, and the non-magnetic Zn$^{2+}$ sites could be occupied by Cu$^{2+}$ with spin-1/2 with the occupancy rate $x$ ranging from $x \simeq0.05$ to $1$.

When $x=1$ and all the interlayer Zn$^{2+}$ sites are occupied by Cu$^{2+}$ ions, the resulting mineral barlowite Cu$_{4}$(OH)$_{6}$FBr (abbreviated as Cu$_{4}$ hereafter) 
undergoes antiferromagnetic  long range order at $T_\text{N} = 15$~K with canted ferromagnetic moments~\cite{Elliott2014,Han2014barlowite,Jeschke2015,SMAHA2018123,Pasco2018barlowite,Ranjith2018,Feng2018barlowite,Tustain2018PRM,Tustain:2020aa,Smaha2020,Smaha2020_PRM,Tustain2021,SmahaRIXSprb}.  See Fig.1 for the crystal structure of the orthorhombic barlowite 1 phase of Cu$_{4}$~\cite{Smaha2020}.   On the other hand, when the occupancy rate is minimized to $x\simeq0.05$, one obtains Zn-barlowite (Zn$_{0.95}$Cu$_{0.05}$)Cu$_3$(OH)$_{6}$FBr (abbreviated as Zn$_{0.95}$ hereafter) that remains paramagnetic down to sub 1K range without long range order~\cite{SMAHA2018123,Smaha2020,Smaha2020_PRM,WangNatPhys2021,WangPRL2022,Yuan2022}.  The defect concentration $x\simeq0.05$ of Zn-barlowite Zn$_{0.95}$~\cite{Smaha2020_PRM} is smaller than that of herbertsmithite (Zn$_{0.85}$Cu$_{0.15}$)Cu$_3$(OH)$_{6}$Cl$_{2}$~\cite{Freedman2010} by a factor of three, hence Zn$_{0.95}$ is considered the least disordered kagome quantum spin liquid candidate material known to date that does not undergo a symmetry breaking magnetic long range order.  Note that YCu$_3$(OH)$_{6}$Cl$_{3}$~\cite{Sun2016} and Y$_3$Cu$_9$(OH)$_{19}$Cl$_{8}$~\cite{Puphal2017} may be free from such inter site defects, but unfortunately these kagome materials undergo long range order at finite temperatures.

\begin{figure}
\centering
\includegraphics[width=3.5in]{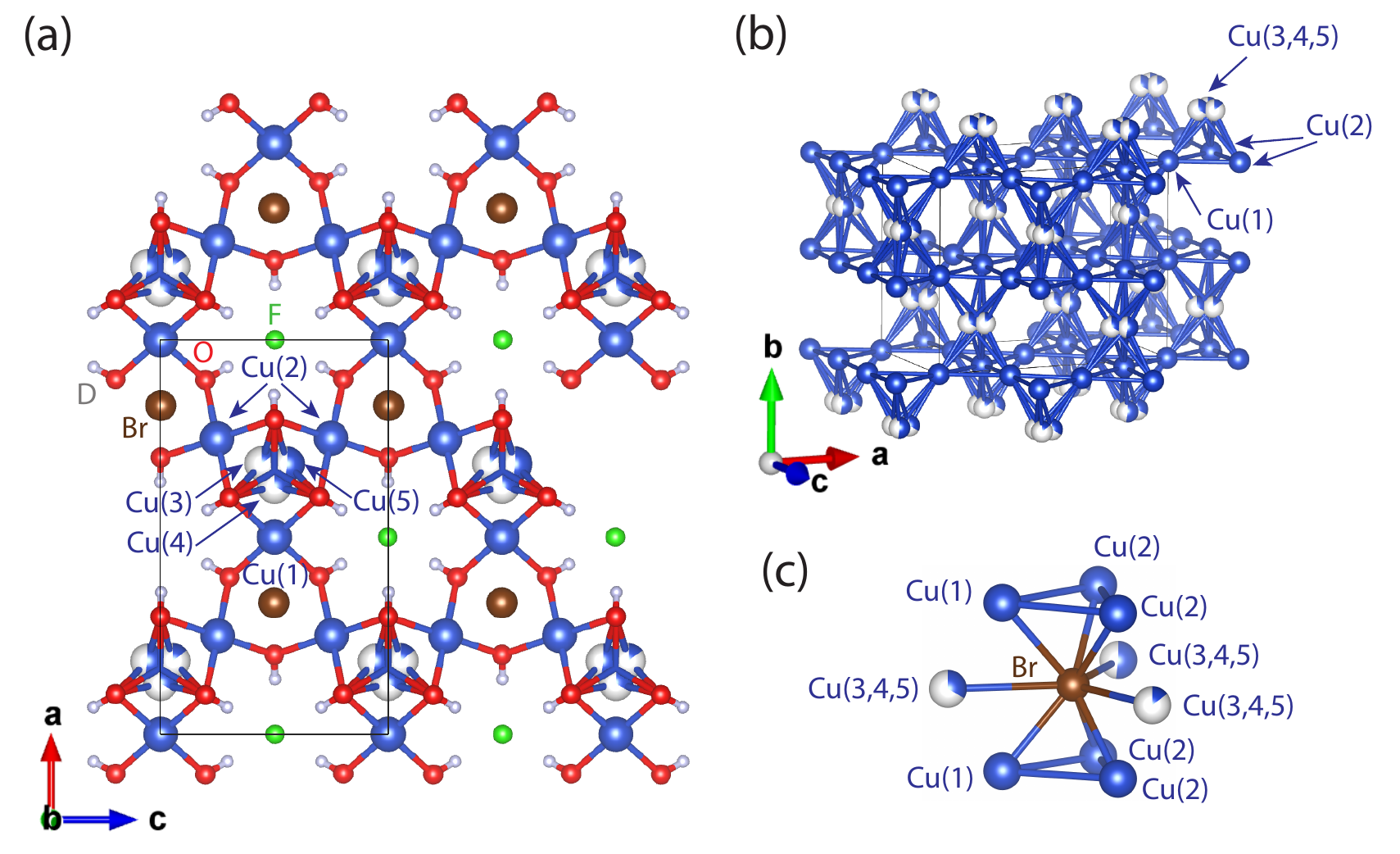}
\caption{(a) The crystal structure of the orthorhombic ``barlowite 1" phase of Cu$_4$(OD)$_6$FBr~\cite{Smaha2020} viewed along the {\it b}-axis, visualized using Vesta~\cite{Momma:2011aa}: Cu (blue), O (red), D (gray), F (green), and Br (brown).  Black line shows the unit cell.  Distorted kagome planes are formed by Cu(1,2), O, and D sites, while Cu(3,4,5), F and Br sites are located between two adjacent kagome planes.  (b) A tilted skeleton view depicting only the Cu sites, showing four interlayers formed by Cu(3,4,5) sites that sandwich three slightly distorted kagome layers.  (c) Local geometry of the Br site surrounded by two Cu(1) and four Cu(2) sites in adjacent kagome planes as well as three Cu(3,4,5) sites within the same interlayer.  For clarity, only Br and Cu sites are shown.
}
\label{structure}
\end{figure}

Although long range order is absent in Zn$_{0.95}$ and herbertsmithite, it has become increasingly clear that the interlayer Cu$^{2+}$ defect spins occupying the Zn$^{2+}$ sites significantly perturb their physical properties, such as the enhanced bulk magnetization at low temperatures.  Moreover, our recent $^{63}$Cu NQR measurements on Zn$_{0.95}$ showed that spin singlets emerge gradually below 30~K with inhomogeneous gaps~\cite{WangNatPhys2021}, and up to $\sim60$\% of the Cu spins within kagome planes are polarized instead by interlayer defect Cu spins at low temperatures~\cite{Yuan2022}.  It remains to be clarified exactly how the interlayer Cu$^{2+}$ defect spins couple with and affect the adjacent kagome planes.  In this context, Cu$_4$ provides a useful avenue to investigate the interplay between the interlayer Cu$^{2+}$ spins and adjacent kagome layers.

The motivation outlined above has led to concerted efforts to clarify the nature of magnetism in Cu$_4$ barlowite, but the complicated structure makes it difficult to reach a consensus; this is in part because the crystallographic and physical properties of Cu$_4$ are known to depend slightly on the synthesis route~\cite{SMAHA2018123,Smaha2020}.  This study focuses solely on the orthorhombic variant of Cu$_4$, barlowite 1 phase identified in~\cite{SMAHA2018123,Smaha2020}.  In barlowite 1, the equilateral triangles of the kagome planes are slightly elongated along the {\it a}-axis below the orthorhombic structural transition at 265~K, resulting in two distinct Cu(1) and Cu(2) sites within the kgome planes as shown in Fig.~1(a).  The interlayer Cu(3,4,5) sites occupy three distinct locations between two adjacent kagome planes near the center of the triangles formed by Cu(1) and two Cu(2) sites; Cu(3) and Cu(5) sites are slightly shifted toward Cu(2) sites, while Cu(4) sites are slightly shifted toward Cu(1) sites.  The overall population ratio between the kagome Cu(1), kagome Cu(2), and the three interlayer Cu(3,4,5) sites is 1 vs. 2 vs. 1.  Among the interlayer Cu(3,4,5) sites, the occupancies of the Cu(3), Cu(4), and Cu(5) sites are 52\%, 33\%, and 15\%~\cite{Smaha2020}.  We note that a different synthesis route produces another variant, barlowite 2 phase of Cu$_4$, which remains hexagonal at low temperature and tends toward a pinwheel valence bond crystal ground state~\cite{Smaha2020}.  

In the earlier studies of Cu$_4$ barlowite, it has been agreed that the interlayer Cu(3,4,5) sites possess a robust ordered moment of $\sim0.5\mu_\text{B}$ or even greater, where $\mu_\text{B}$ represents a Bohr magneton~\cite{Feng2018barlowite,Tustain2018PRM}; it is a typical magnitude for quasi two dimensional Heisenberg antiferromagnet with quantum reduction~\cite{Manousakis1991PRM}.  On the other hand, the average moments within the kagome planes at Cu(1) or Cu(2) sites may be small~\cite{Feng2018barlowite,Tustain2018PRM}.  However, no firm consensus has been reached yet on the unique spin structure established below $T_\text{N}$.  For the barlowite 2 variant, recent work suggests instead that the kagome Cu(1) and Cu(2) sites may host the {\bf q~}={~\bf 0} structure with sizable moments~\cite{Smaha2020}, which is generally favored by the frustrated kagome lattice to accommodate a high degree of degeneracy.  Equally perplexing is that the magnetically ordered state of Cu$_4$ appears to emerge below $T_\text{N}$ through a conventional three-dimensional long range order in a $\mu$SR study~\cite{Tustain:2020aa}, while an early $^{79}$Br NMR work emphasized somewhat unconventional nature of the ordered state that seems slowly freezing across $T_\text{N}$~\cite{Ranjith2018}.   This apparent discrepancy may be caused by different samples, and underscores the importance of working on a phase pure barlowite sample.

In this article, we report a detailed $^{79}$Br and $^{63}$Cu nuclear quadrupole resonance (NQR) investigation of Cu$_4$ in zero external magnetic field using a phase-pure, well-characterized powder sample of orthorhombic barlowite 1~\cite{Smaha2020}, and compare the results with the $^{19}$F NMR data we recently reported for the same sample specimen~\cite{Yuan2022}. The NQR properties of the hexagonal barlowite 2 phase are beyond the scope of the present work, because only a limited amount of barlowite 2 sample can be synthesized in the form of very small single crystals, whose total volume is too small and unsuitable for NQR measurements.   From the $^{79}$Br NQR measurements, we demonstrate that at least one of the crystallographically inequivalent Cu sites, most likely some of the interlayer Cu(3,4,5) sites, exhibit the typical signature of three-dimensional antiferromagnetic long range order, i.e. the critical slowing down of spin fluctuations below $\sim20$~K that precedes the long range order at $T_\text{N}$.  In contrast, the gradual loss of $^{63}$Cu NQR signal intensity suggests that gradual spin freezing sets in below $\sim30$~K within the kagome layers.  These contrasting findings suggest an interesting possibility that the frustrated kagome planes in barlowite 1 are forced into the ordered state when the interlayer Cu(3,4,5) sites undergo a robust long range order.  This is reminiscent of the successive magnetic transitions observed for barlowite 2, where ordering of the interlayer Cu moments upon cooling drives the ordering of the kagome Cu moments~\cite{Smaha2020}.

\begin{figure}
\centering
\includegraphics[width=3.5in]{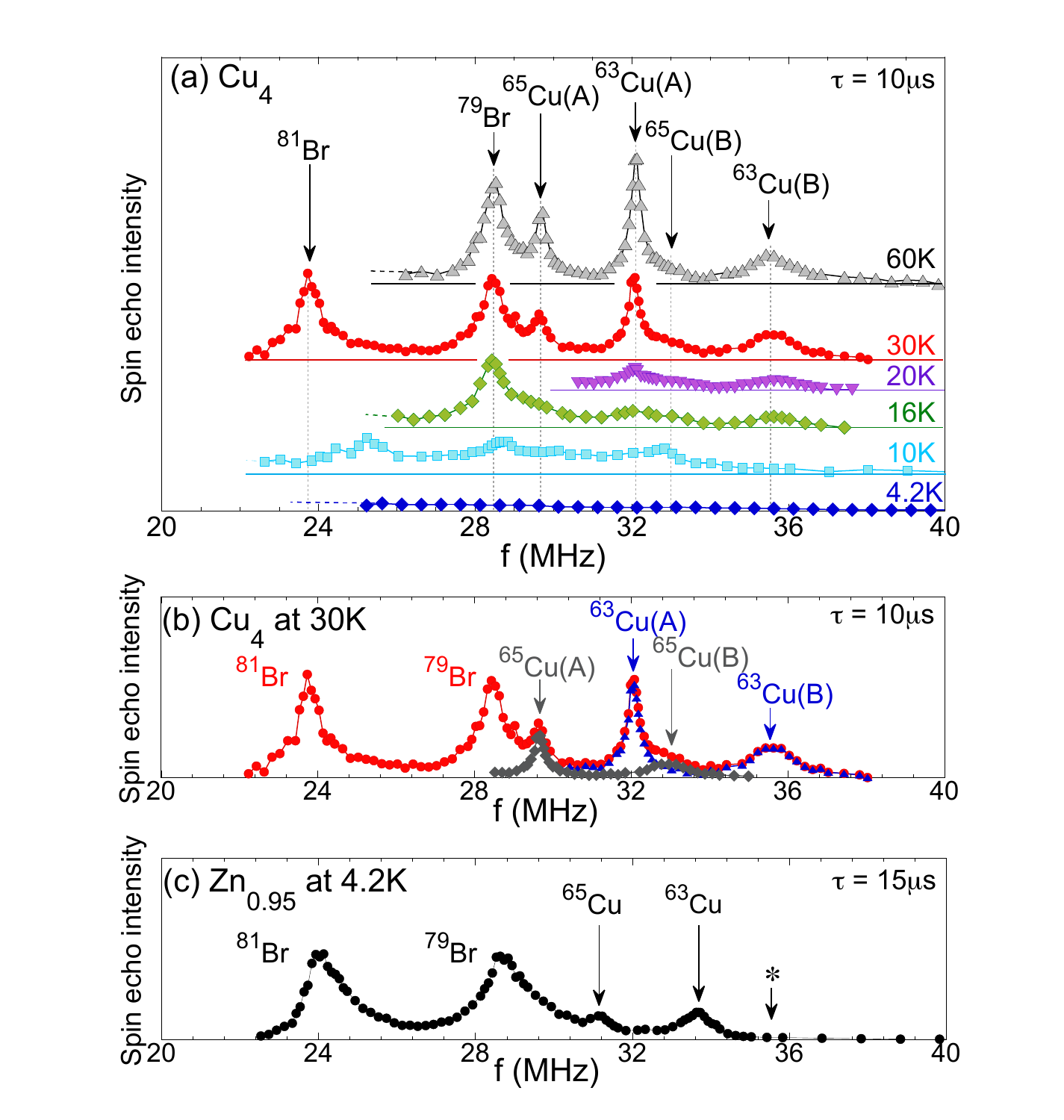}
\caption{(a) $^{79,81}$Br  and $^{63,65}$Cu NQR lineshapes observed for Cu$_4$(OD)$_6$FBr at representative temperatures with the separation time $\tau = 10~\mu$s between the $\pi$/2 and $\pi$ pulses.  The origin of the vertical axis is shifted at different temperatures for clarity.  (b) Deconvolution of the $^{63,65}$Cu NQR lineshape observed at 30~K (\textcolor{red}{$\bullet$}, same data as in panel (a)) into $^{63}$Cu only (\textcolor{blue}{$\blacktriangle$}) and $^{65}$Cu only (\textcolor{gray}{$\filleddiamond$}) NQR lineshapes.  See the main text for the details of the procedures for deconvolution.  (c) $^{79,81}$Br  and $^{63,65}$Cu NQR lineshapes observed for (Zn$_{0.95}$Cu$_{0.05}$)Cu$_3$(OD)$_6$FBr at 4.2~K with somewhat longer $\tau = 15~\mu$s~\cite{WangNatPhys2021}.  The * mark shows the small signals observed at the same frequency as $^{63}$Cu(B) sites in Cu$_4$.  All solid and dashed lines are a guide for eyes.
}
\label{NQRlineshape}
\end{figure}

\begin{figure}
\centering
\includegraphics[width=3.2in]{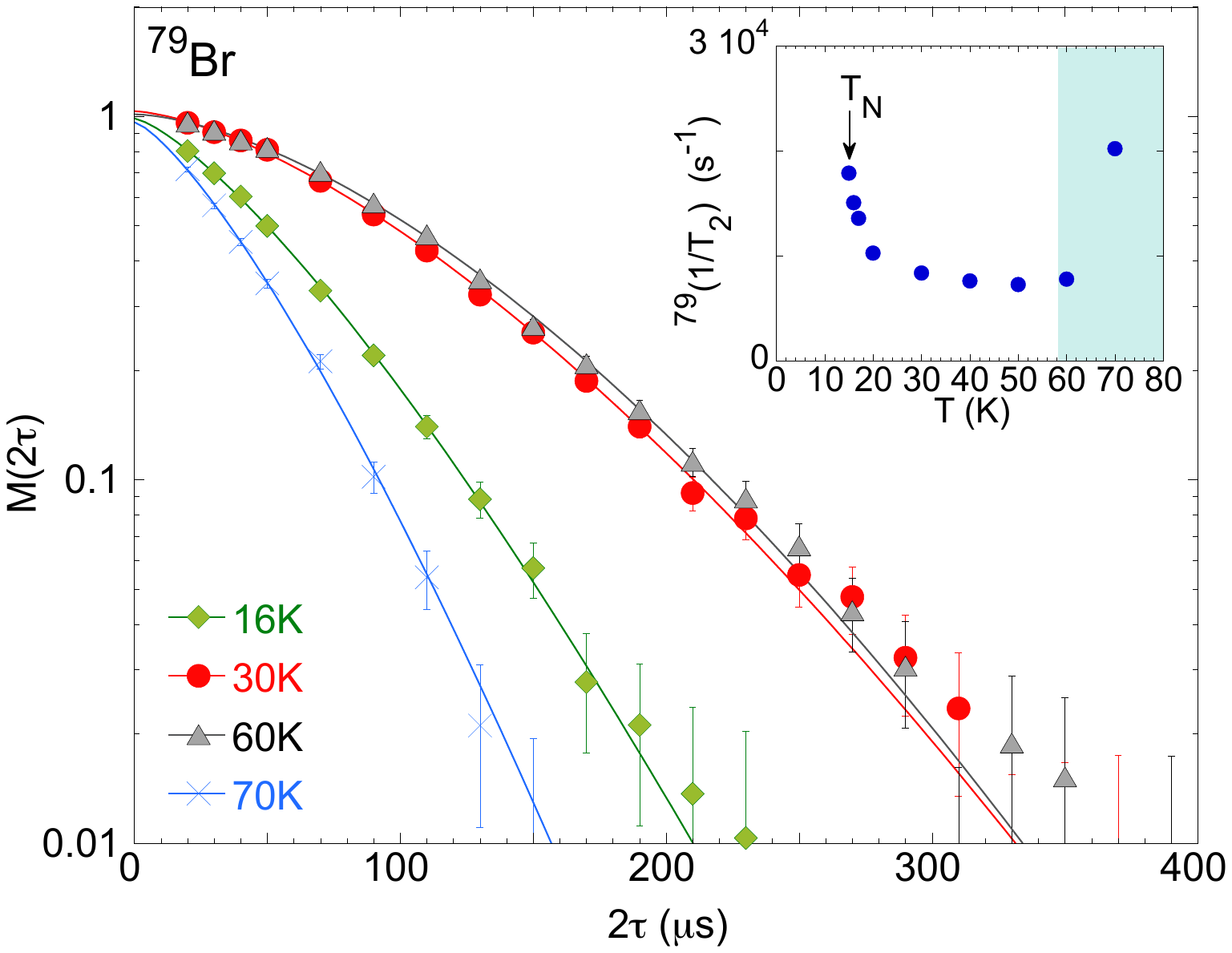}
\caption{(Main panel) The transverse spin echo decay curve $M(2\tau)$ observed for the $^{79}$Br sites of Cu$_4$.  The solid curves are the best fit with an empirical form $M(2\tau) = M(0)~exp(-[2\tau/^{79}T_{2}]^{\alpha})$, where $\alpha$ was consistently found to be $\sim1.3$ from the best fit.  We corrected the intensity for the Boltzmann factor by multiplying temperature $T$, and set the overall intensity by normalizing the 60~K intensity extrapolated to $2\tau = 0$ as 1.  No other normalization was made at different temperatures, but the spin echo intensity in the limit of $2\tau=0$ is conserved at different temperatures above $T_\text{N}$.    (Inset) The transverse relaxation rate $^{79}(1/T_{2})$ observed at $^{79}$Br sites of Cu$_4$.  Throughout this article, the region above $\sim60$~K with light blue shading marks the temperature range where the slow lattice dynamics enhance the NMR relaxation rates of the quadrupolar nuclei~\cite{Ranjith2018,WangPRL2022}.
}
\label{BrT2}
\end{figure}

\section{Experimental}
We synthesized the deuterated powder sample of the barlowite 1 phase (Cu$_4$) by mixing Cu$_2$(OH)$_2$CO$_3$ (Alfa, Cu 55\%), NH$_4$F (Alfa, 96\%), HBr (Alfa, 48\% wt), and 36 mL D$_2$O (Aldrich, 99.9\%) in a 45mL autoclave, which was heated over 3 h to 175 °C and held for 72 h before being cooled to room temperature over 48 h. The products were recovered by filtration and washed with deionized H$_2$O, yielding polycrystalline Cu$_4$~\cite{SMAHA2018123,Smaha2020}.

We use a deuterated sample for NQR measurements instead of a protonated sample for two reasons.  First, we synthesized and used deuterated Cu$_4$ sample for neutron scattering measurements, because protonated sample would suffer from strong background signals in neutron scattering~\cite{Smaha2020}.  Second, the very broad and strong $^{1}$H NMR signals from protonated Cu$_4$ is known to be superposed on $^{19}$F NMR signals~\cite{Ranjith2018}, which would have prevented us from conducting two-dimensional NMR measurements across the whole  $^{19}$F NMR lineshape~\cite{Yuan2022}.  We emphasize that deuteration does not affect other physical properties.

We conducted $^{63,65}$Cu and $^{79,81}$Br (all with nuclear spin 3/2) NQR measurements with standard pulsed NMR spectrometers using the same Cu$_4$ sample  investigated previously with $^{19}$F NMR~\cite{Yuan2022}.  For the measurements of the nuclear spin-lattice relaxation rate $1/T_1$, we used a separation time $\tau = 6~\mu$s between the $\pi$/2 and $\pi$ radio frequency pulses so that we capture as much signals with short transverse relaxation time $T_2$ as possible.  For the NQR lineshape measurements, we used a somewhat longer pulse separation time $\tau = 10~\mu$s to reduce the ring down in the lower frequency range below $\sim$30~MHz.  The typical length for the $\pi$/2 ($\pi$) pulse was 2.8~$\mu$s (5.6~$\mu$s).

We measured the $1/T_1$ recovery curve $M_\text{z}(t)$ by applying an inversion pulse prior to the spin echo sequence, and fitted the observed results with a stretched exponential form, $M_\text{z}(t) = M_\text{0} - A~\text{exp}(-(3t/T_{1})^{\beta})$, where the factor 3 originates from the matrix element for nuclear spin $I=3/2$; besides $1/T_1$, the saturated nuclear magnetization $M_\text{0}$, the inverted nuclear magnetization $A$, and the stretched exponent $\beta$ are the free parameters of the fit.  We found that the distribution in the magnitude of $1/T_1$ is fairly small and  $\beta \simeq 1$ in most of the temperature range of our concern between $T_\text{N}=15$~K and 60~K.  The only exception was one of the two Cu peaks below 30~K, where the glassy nature of spin dynamics strongly manifests itself and the signal intensity is gradually wiped out. 

\section{Results and Discussion}
\subsection{$^{79}$Br NQR lineshapes}
In Fig.~2(a), we summarize the $^{79,81}$Br and $^{63,65}$Cu NQR lineshapes observed at representative temperatures in zero applied magnetic field.  The overall intensity of each lineshape is corrected for the Boltzmann factor by multiplying temperature $T$, and also for the frequency-dependent sensitivity by dividing the observed intensity by the frequency squared, $f^{2}$.  The natural abundance of $^{79}$Br and $^{81}$Br nuclei is 51\% and 49\%, respectively, and nearly equal NQR intensities observed at 30~K for $^{79}$Br and $^{81}$Br confirms that our procedures work well for the estimation of the relative intensities of different peaks.

Since the $^{79,81}$Br NQR peak frequencies are proportional to their nuclear quadrupole moment $^{79,81}Q$, the ratio between the peak frequency of $^{81}$Br and $^{79}$Br is expected to be $^{81}f/^{79}f = ~^{81}Q/^{79}Q = 0.837$.  Likewise, the ratio between the peak frequencies of $^{65}$Cu and $^{63}$Cu would be $^{65}f/^{63}f=~^{65}Q/^{63}Q = 0.927$.  Therefore, we can distinguish the $^{79,81}$Br and and $^{63,65}$Cu peaks as marked in Fig.~2(a).  The observed $^{79,81}$Br peak frequencies are similar to an earlier report~\cite{Ranjith2018}, but our $^{79,81}$Br peaks are well separated from each other; it assures us that our sample of barlowite 1 phase has less structural disorder.  The intensity of $^{65}$Cu peak is smaller than that of $^{63}$Cu peak, because the natural abundance of $^{65}$Cu (31\%) is much smaller than that of $^{63}$Cu (69\%).  

Below $T_\text{N}$, the static hyperfine magnetic fields arising from ordered Cu moments perturb the energy levels of the nuclear spins and shift the resonance frequencies.  At 4.2~K, we confirmed that the Zeeman-perturbed NQR lineshape extends above 60~MHz.  One can sometimes gain local information about the magnitude and orientation of the ordered moments belonging to multiple sub-lattices from comprehensive measurements and analysis of the lineshapes below $T_\text{N}$~\cite{Yoshinari1990Nd2CuO4}.  However, the analysis of the Zeeman-perturbed NQR lineshapes is a highly complicated problem if the hyperfine fields are comparable to NQR frequencies and perturbation analysis is not applicable~\cite{Hunt2001}, which appears to be the case here.   Accordingly, the ordered state below $T_\text{N}$ is beyond the scope of the present work.  

The intensities of the $^{79,81}$Br NQR peaks are conserved below 60~K down to 30~K.  This is consistent with the general expectations for paramagnetic insulators, because the number of nuclear spins detected in our spin echo measurements is conserved.  The intensity of the $^{79}$Br peak at 16~K appears somewhat smaller, simply because the transverse relaxation rate $^{79}(1/T_{2})$ of $^{79}$Br sites is enhanced near $T_\text{N}$, as summarized in the inset of Fig.~3.  Notice that the extrapolation of the transverse nuclear magnetization $M(2\tau)$ to $2\tau=0$ in the main panel of Fig.~3 is actually conserved at 16~K.  We also observed enhancement of $^{79}(1/T_{2})$ above 60~K in the region marked with blue shading in the inset of Fig.~3.  It is due to gradual freezing of the lattice distortion commonly observed for Cu$_4$ \cite{Ranjith2018}, Zn$_{0.95}$ \cite{WangPRL2022} and herbertsmithite (Zn$_{0.85}$Cu$_{0.15}$)Cu$_3$(OH)$_6$Cl$_2$~\cite{Imai2008,Fu2015,Zorko2017}.  

In the case of Zn$_{0.95}$, the spin echo decay curve $M(2\tau)$ at $^{79}$Br sites develops a highly damped oscillation due to indirect nuclear spin-spin coupling effect below 50~K \cite{WangPRL2022}.  It indicated Br-Br  dimer formation for up to $\sim50$\% of Br sites, as spin singlets gradually emerge in the kagome planes with inhomogeneous gaps \cite{WangNatPhys2021}.  Such oscillatory behaviors in $M(2\tau)$ were previously observed for other spin singlet materials as well \cite{Kodama2002,Kikuchi2010,Kikuchi2013}, but we find no such oscillations in the present case. 

\subsection{$^{63}$Cu NQR lineshapes}

Let us now turn our attention to two sets of $^{63,65}$Cu NQR peaks in Fig.~2(a).  In Fig.~2(b), we use the following standard procedures to deconvolute the $^{63,65}$Cu NQR lineshapes observed at 30~K into $^{63}$Cu only and $^{65}$Cu only NQR lineshapes:  First, from the $^{63}$Cu NQR signals observed at $^{63}f \simeq 37$~MHz near the upper end of the frequency range, we estimate the corresponding $^{65}$Cu NQR frequency $^{65}f = (^{65}Q/^{63}Q)\times^{63}f$ ($\simeq 34.3$~MHz).  We also estimate the intensity of the $^{65}$Cu NQR signal at $^{65}f$ from the isotope ratio, $0.31/0.69 = 0.45$.  Then one can plot the deconvoluted $^{65}$Cu NQR lineshape around $\sim34.3$~MHz, as shown with gray diamonds.  We can repeat this procedure down to $^{63}f \simeq 35$~MHz, below which we need to subtract the estimated $^{65}$Cu NQR intensity from the overall intensity to estimate the $^{63}$Cu contribution shown with blue triangles, then use the latter to estimate the corresponding $^{65}$Cu NQR signal at lower frequency.

By repeating these procedures down to $\sim29$~MHz, we identified two distinct $^{63}$Cu peaks in Fig.~2(b): $^{63}$Cu(A) peak at $\sim$32~MHz and $^{63}$Cu(B) peak at $\sim$35.5~MHz.  The bare integrated intensity ratio between the narrow 32~MHz peak and broad 35.5~MHz peak is $1\pm0.15$.  Once we take into account the slightly faster transverse relaxation rate $^{63}(1/T_{2})$ of the 32~MHz peak as shown in Fig.4(c), we estimate the intensity ratio to be $1.3\pm0.2$.   Recall that there are five non-equivalent Cu sites in Cu$_4$, although the differences are small between Cu(1) and Cu(2) sites as well as between Cu(3), Cu(4), and Cu(5) sites.  There are two potential scenarios for site assignments of the $^{63}$Cu(A) and $^{63}$Cu(B) peaks:

Scenario 1.  Note that the $^{63}$Cu(A) and $^{63}$Cu(B) peak frequencies are comparable to the main peak frequency $\sim$33.7~MHz observed for the kagome Cu sites of Zn$_{0.95}$ (see Fig.~2(c) for comparison).  Elongation of the kagome plane along the a-axis in Cu$_4$ generates two inequivalent Cu(1) and Cu(2) sites within the kagome plane, and the latter is twice more abundant than the former.  Since the NQR frequency is generally highly sensitive to the local structural environment, the most plausible scenario is that $^{63}$Cu(A) and $^{63}$Cu(B) peaks arise from the kagome Cu(2) and Cu(1) sites, respectively.  In this scenario, the $^{63}$Cu NQR signals of interlayer Cu(3,4,5) sites are missing in our lineshapes.  That is not surprising in view of the fact that the $^{63}$Cu NQR signals of the interlayer Cu sites in herbertsmithite are not observable despite as much as 15\% occupancy rate~\cite{Imai2008,WangNatPhys2021}.  Moreover, the bulk magnetization of Cu$_4$ as well as Zn$_{0.95}$ and herbertsmithite is very large at low temperatures~\cite{Smaha2020} due primarily to the weakly interacting interlayer defect spins~\cite{Imai2011}.  Since the NMR relaxation rates $1/T_1$ and $1/T_2$ are generally inversely proportional to the local interaction energy scale $J'$ (i.e. $1/T_{1}$, $1/T_{2}\propto 1/J'$~\cite{Moriya1956II}), we expect that the fast relaxation rates at the interlayer Cu(3,4,5) sites wipe out their $^{63}$Cu NQR signals.

The observed intensity ratio 1.3 vs.~1 between Cu(2) and Cu(1) sites is significantly smaller than 2 vs.~1 expected from the crystal structure, and suggests that some of the $^{63}$Cu NQR signals are wiped out within the kagome planes, too, due to their fast relaxation rates.  This may be understandable if we recall that the interlayer Cu$^{2+}$ defect spins can form a cluster with Cu and O sites located in the adjacent kagome layers and induce spin polarization to the latter, as demonstrated earlier in our $^{2}$D~\cite{Imai2011} and $^{17}$O~\cite{Fu2015} single crystal NMR in herbertsmithite and $^{19}$F NMR in Zn$_{0.95}$~\cite{Yuan2022}.  Since the Cu(3) and Cu(5) sites, which account for 67\% of the overall interlayer Cu sites, are shifted from the center of the triangle toward Cu(2) sites~\cite{Smaha2020}, perhaps the small coupling $J'$ between the Cu(2) sites and the Cu(3) and/or Cu(5) sites may be locally enhancing the relaxation rate at the former, thus making some of the Cu(2) sites not observable.

Scenario 2.  On the other hand, as shown in Fig.~5(c), the magnitude of $^{63}(1/T_{1})$ observed at Cu(B) sites is by a factor of $\sim 3$ greater than at Cu(A) sites.  Moreover, the Cu(B) peak frequency is nearly identical with that of the small signals observed for Zn$_{0.95}$, in which only $\sim$5\% of the interlayer Zn sites are occupied by Cu (see * in Fig.2(c)).  Therefore, an alternate scenario is that the Cu(A) peak arises from the combination of some of the Cu(1) and Cu(2) sites, whereas the Cu(B) peak arises from some of the Cu(3,4,5) sites.  However, given that the combined abundance of the Cu(1) and Cu(2) sites is three times greater that of Cu(3,4,5) sites, this alternate scenario would imply that a majority of the $^{63}$Cu NQR signals from the kagome Cu(1,2) sites are wiped out despite the fairly large exchange coupling $J\simeq 160$~K between the kagome Cu sites that should suppress the relaxation rates.  Accordingly, scenario 2 seems unlikely.   It is also worth recalling that the $^{63}$Cu NQR signals of the kagome Cu sites are readily observable in both herbertsmithite and Zn$_{0.95}$, whereas the interlayer Cu sites are not observable in herbertsmithite despite the greater occupancy rate of 15\%.

\begin{figure}
\centering
\includegraphics[width=3.2in]{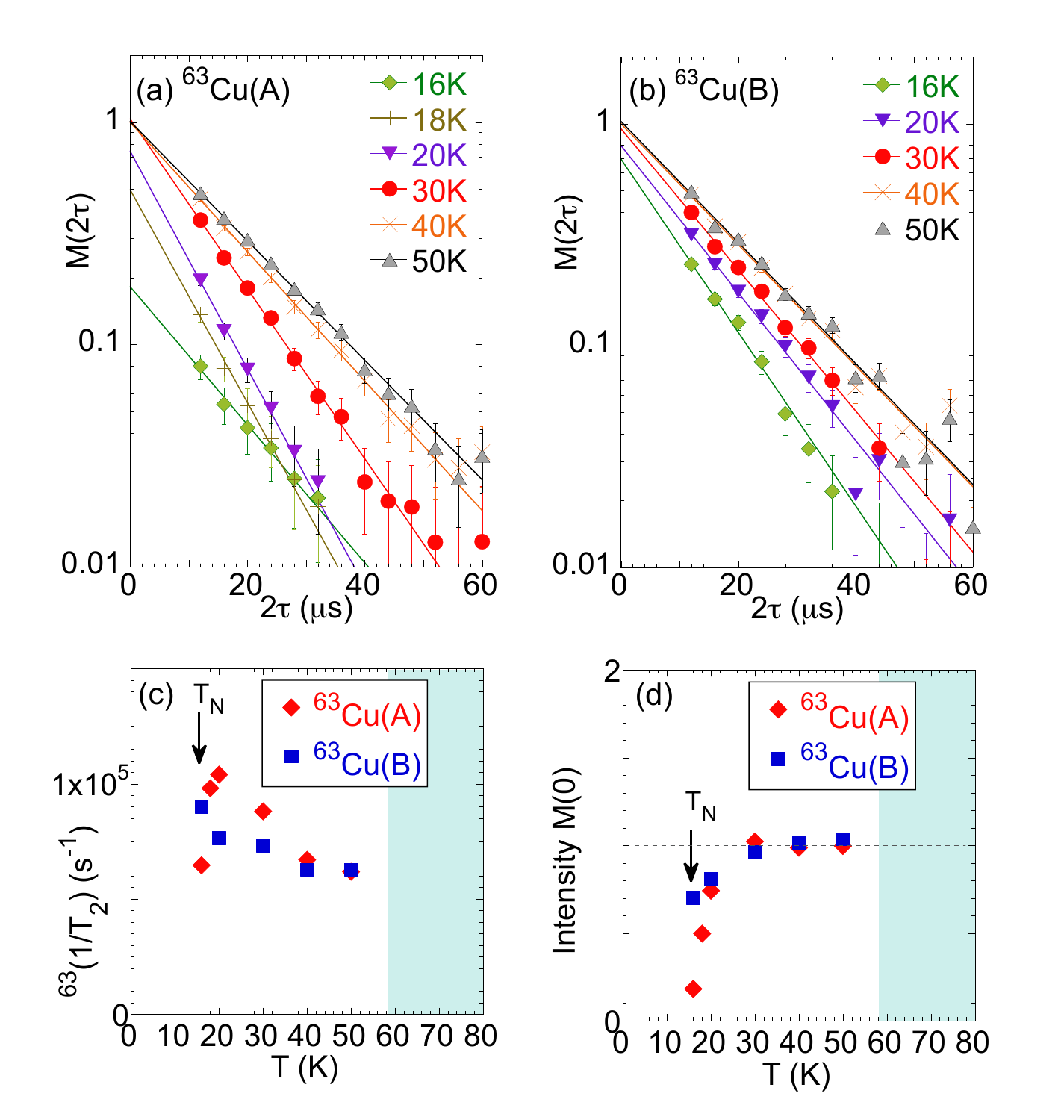}
\caption{The spin echo decay curve $M(2\tau)$ observed for Cu$_4$ at (a) $^{63}$Cu(A) sites at $\sim32$~MHz and (b) $^{63}$Cu(B) sites at $\sim35.5$~MHz.  The spin echo intensity is corrected for the Boltzmann factor by multiplying temperature $T$, and normalized for the intensity observed at 50~K.  The straight lines are the best exponential fit with $M(2\tau) = M(2\tau=0) \times \text{exp}(-2\tau/^{63}T_{2})$.  Notice that the extrapolation of the fit to $2\tau = 0$ is not conserved below 30~K.  (c) The transverse relaxation rate $^{63}(1/T_{2})$ observed at $^{63}$Cu sites, as determined from the exponential fit in (a) and (b).  (d) The signal intensity $M(2\tau=0)$, corrected for both the Boltzmann factor and the transverse relaxation effects.
}
\label{CuT2}
\end{figure}

\subsection{Cu spin dynamics probed by $^{79}$Br NQR}

In Fig.~5(a), we summarize the temperature dependence of $^{79}(1/T_{1})$ measured at the $^{79}$Br NQR peak of Cu$_4$ in comparison to the results observed for Zn$_{0.95}$ \cite{WangNatPhys2021}.  The interlayer $^{79}$Br sites of barlowite are surrounded by two Cu(1) and four Cu(2) sites in two adjacent kagome planes and three Cu(3,4,5) sites within the same interlayer, as shown in Fig.~1(c).  In view of the spatially extended nature of the atomic orbitals of $^{79}$Br, the hyperfine couplings with all these Cu sites probably contribute significantly.   Therefore, $^{79}(1/T_{1})$ is likely to probe low frequency Cu spin dynamics at $\sim28.6$~MHz at all the Cu sites within the kagome layers and interlayers.

The gradual increase of $^{79}(1/T_{1})$ observed for Cu$_4$ below $\sim60$~K is typical when short range spin correlation grows.  This is consistent with the fact that the $^{79}$Br spin echo decay curve of Cu$_4$ does not exhibit the damped oscillation often seen when spin dimers form~\cite{WangPRL2022}.  Below $\sim20$~K,  $^{79}(1/T_{1})$ in Fig.~5(a) as well as $^{79}(1/T_{2})$ in the inset of Fig.~2 quickly grow toward $T_\text{N}$. These divergent behaviors of $^{79}(1/T_{1})$ and $^{79}(1/T_{2})$ slightly above  $T_\text{N}$ are the prototypical signature of critical slowing down of spin fluctuations toward the three-dimensional antiferromagnetic long range order in quasi-two dimensional antiferromagnets~\cite{Borsa1992,NingPRB2014}.  In other words, as far as the $^{79}(1/T_{1})$ and $^{79}(1/T_{2})$ results are concerned, Cu$_4$ appears to be a prototypical three dimensional antiferromagnet.  Also note that the stretched exponent at $^{79}$Br sites hardly deviates from $\beta = 1$ down to $T_\text{N}$ as shown in Fig.~5(d), implying that $^{79}(1/T_{1})$ is not significantly distributed.  Therefore, the $^{79}(1/T_{1})$ results show nearly homogeneous slowing down of {\it at least one} of the Cu sites, but not necessarily all Cu sites.  For example, as long as the hyperfine coupling between the $^{79}$Br nuclear spins and the interlayer Cu(3,4,5) sites is not negligibly small, $^{79}(1/T_{1})$ and $^{79}(1/T_{2})$ would show the signature of critical slowing down toward $T_\text{N}$ when the interlayer Cu(3,4,5) sites undergo a conventional antiferromagnetic long range order, even if Cu(1) and Cu(2) sites do not enter a long range ordered state and remain paramagnetic.  

\begin{figure}
\centering
\includegraphics[width=3.2in]{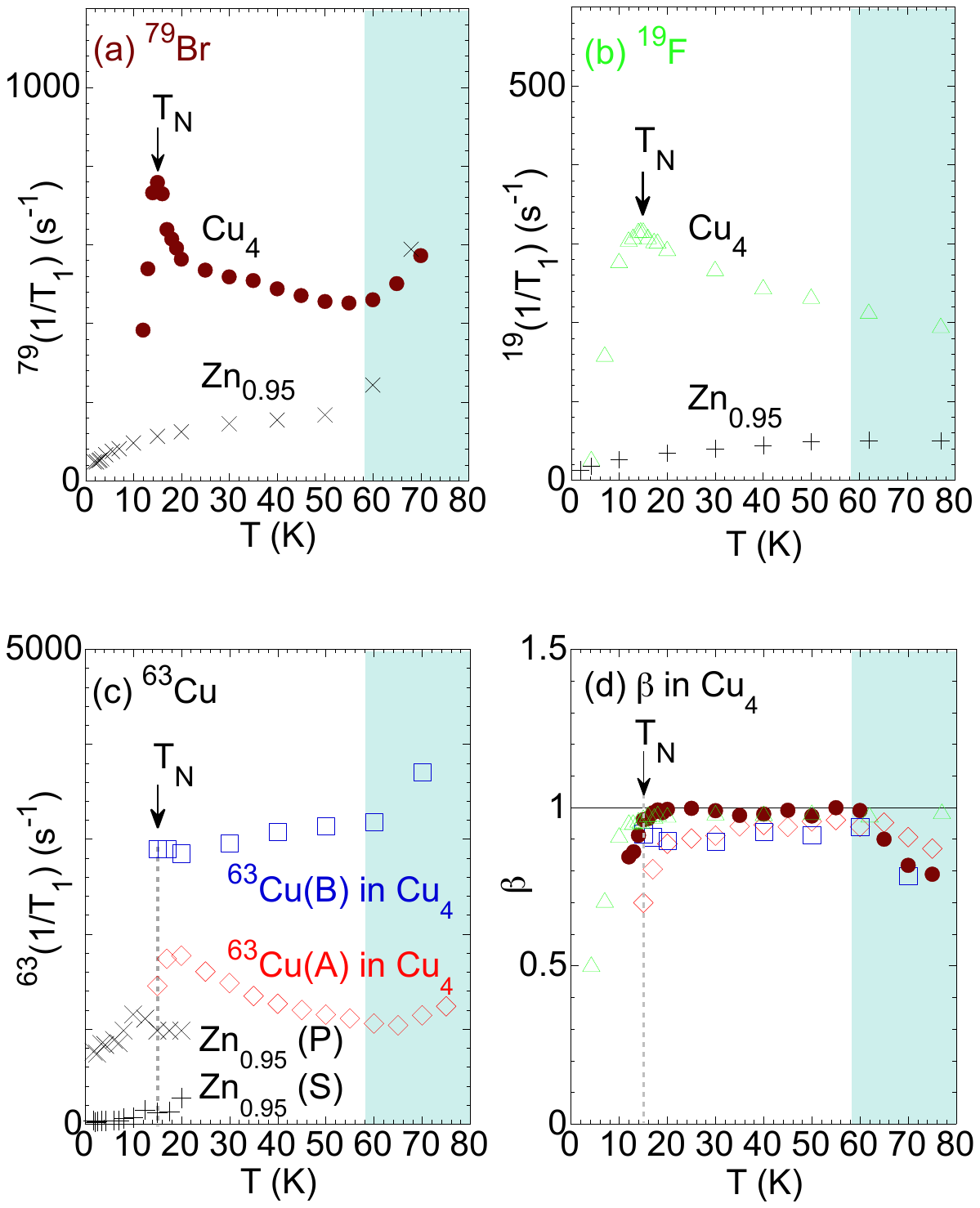}
\caption{(a) $^{79}(1/T_{1})$  observed at $^{79}$Br sites of Cu$_4$ (\textcolor{brown}{$\bullet$}) and Zn$_{0.95}$ ($\times$, from \cite{WangNatPhys2021}).  (b) $^{19}(1/T_{1})$ observed for $^{19}$F sites of Cu$_4$ (\textcolor{green}{$\triangle$}) and Zn$_{0.95}$ (+) in 2.4~T, both from \cite{Yuan2022}.  Note that the upturn of $^{19}(1/T_{1})$ caused by critical slowing down is absent below 20~K.  Since $^{19}$F nuclear spin is 1/2 and hence immune from the fluctuations of the electric field gradient, $^{19}(1/T_{1})$ does not show an upturn above 60~K, either~\cite{WangNatPhys2021}.    (c) $^{63}(1/T_{1})$ observed at $^{63}$Cu(A) (\textcolor{red}{$\diamond$})  and $^{63}$Cu(B) (\textcolor{blue}{$\square$}) peaks of Cu$_4$.  The data points below 30~K should be considered the lower bound of the highly distributed $^{63}(1/T_{1})$ throughout the sample volume, because the $^{63}$Cu sites with larger relaxation rates do not contribute to the spin echo signals detected with a finite delay time $\tau=6~\mu$s.  For comparison, we also show $^{63}(1/T_{1})$ at the paramagnetic Cu site (P)  of Zn$_{0.95}$ (x) and spin singlet Cu site (S) of Zn$_{0.95}$ (+)  as determined by inverse Laplace transform (ILT) $T_1$ analysis of the recovery curve~\cite{WangNatPhys2021}.  (d) The stretched exponent $\beta$ observed for Cu$_4$ at $^{79}$Br (\textcolor{brown}{$\bullet$}), $^{19}$F (\textcolor{green}{$\triangle$}), $^{63}$Cu(A) (\textcolor{red}{$\diamond$})  and $^{63}$Cu(B) (\textcolor{blue}{$\square$}) sites, shown using the same symbols as in panel (a) -  (c). 
}
\label{BrFT1beta}
\end{figure}

It is also interesting to note that our $^{79}(1/T_{1})$ results are noticeably different from the high field NMR results reported in an earlier work~\cite{Ranjith2018}.  The magnitude of our $^{79}(1/T_{1})$ is larger by as much as a factor of $\sim3$, and the enhancement toward  $T_\text{N}$ seems more pronounced than in the previous work.  As noted above, the physical and structural properties of Cu$_4$ are known to be sensitive to synthesis conditions~\cite{Pasco2018barlowite,SMAHA2018123,Smaha2020}, which might explain the apparent discrepancies.  We do note that  $^{79,81}$Br NQR peaks are barely distinguishable in the earlier report~\cite{Ranjith2018}, which indicates a higher level of structural disorder in the sample used for the previous report.  

\subsection{Comparison with $^{19}$F sites}

In Fig.~5(b), we reproduce $^{19}(1/T_{1})$ measured in 2.4~T at the $^{19}$F sites of Cu$_4$ and Zn$_{0.95}$ \cite{Yuan2022}.  $^{19}(1/T_{1})$  of Cu$_4$ gradually grows from 250~K~\cite{Yuan2022}, but exhibits only a broad hump centered at $T_\text{N}$ with no hint of critical slowing down below $\sim20$~K.  To understand this apparently contradicting results, it is useful to recall that the atomic orbitals of $^{19}$F sites are spatially less extended than those of $^{79}$Br.  Therefore, the hyperfine couplings of $^{19}$F sites may be limited to the six nearest neighbor O sites located at a distance of $\sim2.7$~$\text{\AA}$ in each of the two adjacent kagome layers, which in turn bond with Cu(1) and Cu(2) kagome sites.  (The distance between the $^{19}$F  and closest interlayer Cu(3,4,5) sites within the same interlayer is much greater, 3.6~$\text{\AA}$, and no O$^{2-}$ sites exist in between them to transfer the spin from the Cu(3,4,5) sites to $^{19}$F sites.)  In this scenario, if critical slowing down is setting in below 20~K only at the interlayer Cu(3,4,5)  sites, $^{19}(1/T_{1})$ may not show a sharp peak at $T_\text{N}$.  

An alternate scenario to account for the absence of the sharp peak in $^{19}(1/T_{1})$ at $T_\text{N}$ would rely on a complete geometrical cancellation of the transferred hyperfine magnetic fields at $^{19}$F sites due to the wave-vector dependence in the hyperfine form factor~\cite{Shastry1989}.  If all the Cu sites form a commensurate antiferromagnetic structure, in principle, the fluctuating transferred hyperfine fields at the high symmetry $^{19}$F sites can cancel out each other and $1/T_1$ may not  exhibit any sharp anomaly at $T_\text{N}$, as previously demonstrated for the case of square-lattice Heisenberg antiferromagnet~\cite{Thurber1997}.  We cannot logically rule out such a scenario in the present case.  However, as shown in the next subsection, $^{63}$Cu NQR results show hints of spin freezing below $\sim30$~K at one or more Cu sites rather than critical slowing down.

\subsection{$^{63}$Cu signal intensity wipeout, $^{63}(1/T_{1})$, $^{63}(1/T_{2})$}

In Fig.~5(c), we summarize the temperature dependence of $^{63}(1/T_{1})$.  $^{63}(1/T_{1})$ at the Cu(A) peak shows qualitatively similar temperature dependence as $^{79}(1/T_{1})$ above 20~K.  Near $T_\text{N}$, however, $^{63}(1/T_{1})$ at $^{63}$Cu(A) peak as well as $^{63}(1/T_{2})$ shown in Fig.~4(c) appear to decrease toward $T_\text{N}$.  This apparent lack of the signature of antiferromagnetic order, however, does not necessarily mean that Cu spins at Cu(A) peak do not order below $T_\text{N}$.  We note that the $^{63}$Cu(A) NQR peak progressively disappears below 30~K, as shown in Fig.2(a).  Even if we take into account the transverse relaxation effect by extrapolating the spin echo decay curves in Fig.~4(a) to $2\tau =0$, the extrapolated intensity $M(2\tau =0)$ decreases dramatically toward 16~K as summarized in Fig.~4(d).  

This anomaly implies that Cu(A) sites develop a large distribution in the transverse relaxation time $T_2$ below 30~K, and the signal detection becomes impossible for some nuclei when their $T_2$ becomes shorter than the finite pulse separation time $\tau=6$~$\mu$s. Analogous signal intensity loss without any apparent anomalies in the relaxation rates for the {\it observable parts of the signals} is often seen when spin freezing sets in inhomogeneously with highly distributed $^{63}(1/T_{1})$ and $^{63}(1/T_{2})$; examples include slow spin fluctuations in charge ordered high $T_{c}$ cuprates~\cite{Hunt2001,Imai2017,Arsenault2020}, and incipient spin freezing in Kitaev lattice Cu$_2$IrO$_3$~\cite{Takahashi2019} as well as herbertsmithite and Zn$_{0.95}$~\cite{WangNatPhys2021}.

These considerations also indicate that $^{63}(1/T_{1})$ and $^{63}(1/T_{2})$ plotted below 30~K reflect only the Cu(A) sites which are yet to begin to freeze, and hence the data points of $^{63}(1/T_{1})$ and $^{63}(1/T_{2})$ below 30~K should be considered only the lower bound of their distribution in space.  Besides the signal intensity loss, the deviation of the stretched exponent $\beta$ from 1 near $T_\text{N}$ in Fig.~5(d) underscores the inhomogeneous nature of spin fluctuations at Cu(A) sites as well.    We also confirmed from the inverse Laplace transform analysis of the $1/T_1$ recovery curve that unlike the case of Zn$_{0.95}$~\cite{WangNatPhys2021} no split off peak of spin singlets emerges in the density distribution function $P(1/T_{1})$ when the distribution in $^{63}(1/T_{1})$ grows.   

The signal intensity loss at Cu(B) peak toward $T_\text{N}$ is far less pronounced than at Cu(A) peak, as shown in Fig.~4(d).  The apparent reduction of the Cu(B) peak intensity toward $T_\text{N}$ in Fig.~2(a) is primarily because of the enhancement  of $^{63}(1/T_{2})$, as seen in Fig.~4(b) and (c).  This means that most if not all Cu(B) sites continue to contribute to the measurements of $^{63}(1/T_{1})$ and $^{63}(1/T_{2})$.  Indeed $^{63}(1/T_{2})$ at the Cu(B) peak in Fig.~4(c) exhibits a mild divergent behavior toward $T_\text{N}$, although $^{63}(1/T_{1})$ increases only slightly in the same temperature range.  Thus, the signature of spin freezing at the Cu(B) sites is less pronounced.
\section{Summary and Conclusions}
We have reported a comprehensive $^{79}$Br and $^{63}$Cu NQR investigation of antiferromagnetic orthorhombic barlowite 1 phase of Cu$_4$ above $T_\text{N}=15$~K, and compared the results with our earlier $^{19}$F NMR results.  The results of $^{79}(1/T_{1})$ and $^{79}(1/T_{2})$ at $^{79}$Br sites should be considered reflective of the spin dynamics at all the structurally inequivalent Cu sites.  Their divergent behaviors below 20~K toward $T_\text{N}$ indicate that at least one of the inequivalent Cu sites enters the long range ordered state in a fairly conventional manner in the sense that critical slowing down of low frequency Cu spin fluctuations precedes the phase transition.

On the other hand, $^{19}(1/T_{1})$ at $^{19}$F sites, which is more likely to probe only the kagome Cu(1,2) sites due to their proximity, shows only a broad hump at $T_\text{N}$ without any hint of critical slowing down below 20~K.  Two sets of observable Cu NQR signals show little evidence for critical slowing down, either.  Instead, the Cu NQR signal intensity is gradually wiped out below 30~K, which is the typical signature of inhomogeneous spin freezing.  Combining all the observations, the most likely scenario is: (a) Cu(A) and Cu(B) peaks arise from Cu(2) and Cu(1) sites within the kagome planes, (b) it is the Cu NQR signals at Cu(3,4,5) that are missing in the lineshape in Fig.~2, as explained in detail in Scenario 1 in section III B, (c) Cu(3,4,5) sites enter the long range ordered state in a rather conventional manner, and hence $^{79}$Br sites exhibit the signature of critical slowing down, but (d) kagome Cu(1) and Cu(2) sites enter the ordered state in a glassy manner, resulting in the $^{63}$Cu NQR signal intensity loss below 30~K.

The contrasting behaviors inferred for the interlayer and kagome layers suggest that magnetic frustration effects within the kagome layers play significant roles in the spin dynamics of the barlowite 1 phase of Cu$_4$, even though elongation of the kagome layers along the {\it a}-axis partially alleviates the frustration effects.   Although Cu$_4$ is a three-dimensional antiferromagnet, the spin-1/2 kagome planes are in close proximity to other competing ground states, such as valence bond crystals and quantum spin liquids. Here, the robust ordering of the interlayer spins may be the deciding factor in inducing the ordering of the kagome moments.  This scenario is consistent with the earlier finding that the single crystal sample of barlowite 2 phase with slightly different crystal structure exhibits a potential signature of the pinwheel valence bond phase below 10~K due to frustration effects, before the kagome planes enter the ${\bf q} = {\bf 0}$ long range ordered state at $T_\text{N}=6$~K, induced by ordering of the interlayer moments~\cite{Smaha2020}.  It would be interesting to compare the NQR results between barlowite 1 and barlowite 2 phases, but the latter has been successfully synthesized only by a minute amount unsuitable for NQR measurements~\cite{Smaha2020}.

\begin{acknowledgments}
The work at McMaster was supported by NSERC (T.I.).  The work at Stanford and SLAC (sample synthesis and characterization) was supported by the U.S. Department of Energy (DOE), Office of Science, Basic Energy Sciences, Materials Sciences and Engineering Division, under contract no. DE-AC02-76SF00515 (Y.S.L. and J.W.).  R.W.S. was supported by a NSF Graduate Research Fellowship (DGE-1656518).  T.I. acknowledges that the manuscript was finalized at Kavli Institute for Theoretical Physics, University of California at Santa Barbara, during his participation in the qmagnets23 program.
\end{acknowledgments}

\bibliographystyle{apsrev4-2}


%



\end{document}